# Solitons in higher-order topological insulator created by unit cell twisting

Yaroslav V. Kartashov

*Institute of Spectroscopy, Russian Academy of Sciences, 108840, Troitsk, Moscow, Russia*
*kartashov@isan.troitsk.ru*

We show that higher-order topological insulators can be created from usual square structure by twisting waveguides in each unit cell around the axis passing through the center of the unit cell, even without changing intracell distance between waveguides. When applied to usual square array, this approach produces two-dimensional generalization of Su-Schrieffer-Heeger (SSH) structure supporting topological corner modes with propagation constants belonging to two forbidden spectral gaps opening only for twist angles from certain interval. In contrast to usual SSH arrays, where higher-order topology is typically introduced by diagonal waveguide shifts and only one type of corner states exists, our SSH-like structure in topological phase supports two co-existing types of in-phase and out-of-phase corner modes appearing in two different topological gaps that open in the spectrum. Therefore, twisting of the unit cell qualitatively changes topological properties of the system, offering a new degree of freedom in creation of higher-order topological phases. In material with focusing cubic nonlinearity two coexisting types of topological corner solitons emerge from these modes, whose existence and stability properties are studied here. Despite different internal structure, both such modes can be simultaneously dynamically stable in the appreciable part of the topological gap.

Photonic systems that may possess sufficiently strong nonlinear response offer a unique testbed for the exploration of rich interplay between nontrivial topology of spectral bands and nonlinearity of the medium. While nontrivial topology of the system is typically associated with the emergence of topologically protected in-gap edge states (see reviews [1-3] and references therein), nonlinearity adds required tunability into topological systems enabling control over localization degree, shapes, propagation paths and velocity of the topological edge states, see recent reviews [4-7]. Self-action of light in topological systems may result in the formation of topological solitons – unique states that bifurcate from linear topological modes residing in spectral gaps and inherit topological protection of such modes and their resistance to disorder. In some systems, including Chern insulators, topological solitons may even inherit the ability to travel along the edge of topological structure while maintaining or periodically restoring their localized profiles (in some cases the system may be even driven by nonlinearity into topological phase, where light tends to travel along the edge with reduced diffractive broadening [8-10]). Different types of topological solitons were predicted and observed to date, including self-sustained states in the bulk of periodically modulated periodic lattices [11, 12], travelling edge solitons [13-15] in waveguide-based Floquet insulators [16], polaritonic [17-19] and atomic [20] systems. Particular attention has been devoted to the development of theory of travelling edge solitons in discrete [21-24] and continuous [25-28] topological systems based on helical waveguide arrays, as well as in various valley Hall structures [29-31]. Immobile topological solitons have been also observed in one-dimensional topological chains, akin to SSH lattices [32], where nontrivial topological phases usually appear due to controllable shift of sites within unit cell of the lattice [33-42]. It should be mentioned that periodically modulated SSH lattices support specific type of periodically oscillating topological edge solitons emerging from $\pi$-modes [43,44] that were recently observed experimentally [45].

Another intriguing class of topological systems, where investigation of nonlinear effects has just started, is represented by higher-order topological insulators (HOTIs). The most distinctive feature of HOTI is that it can support topological states, whose dimensionality is at least by 2 lower than dimensionality of delocalized bulk modes (see recent review [46]). For example, in two-dimensional photonic HOTIs one can observe the formation of effectively zero-dimensional corner states that frequently coexist with effectively one-dimensional edge states. The common approach to realization of HOTI is based on shift of sites within unit cell of the structure that simultaneously affects inter- and intracell spacing and coupling strengths. Linear photonic HOTIs created with this approach have been have been demonstrated using structures with different symmetries, including square, kagome, or honeycomb ones [47-54]. Such HOTIs were also constructed on aperiodic lattices with disclinations and bulk defects [55-57], see review [58]. Topological solitons in HOTIs have been observed in [59], while nonlinear control of bound states in continuum in such insulations has been reported in [60], see also realizations in polariton [61,62] and atomic [63] condensates, and in general tight-binding models [64]. Such systems can support corner and hinge light bullets without energy threshold [65,66]. A variety of self-sustained states is substantially extended in HOTIs with disclinations [67,68] that can support rich families of topological vortex solitons [69] and in fractal structures [70]. Notice that Floquet version of nonlinear HOTI and solitons in it have been realized very recently in [45].

The properties of higher-order topological modes, and hence of solitons emanating from them, qualitatively depend on the method of construction of HOTI. For example, usual shift of sites towards the periphery/center of the unit cell commonly employed in the literature imposes certain symmetry on the emerging corner states and determines the structure of topological gap appearing in the spectrum of the system. Such shifts rarely lead to the appearance of states with distinct symmetries from different topological gaps, although the exceptions have been reported in some linear systems [71]. Here I show that higher-order topological phase can be induced in periodic arrays by twisting waveguides in the unit cell with respect to the axis passing through the center of the cell. As a particular example, I consider square waveguide array that transforms after such twist into SSH-like insulator supporting two different types of corner states, with different phase structure, that give rise to corner solitons with distinct stability properties in the presence of nonlinearity.

It is important to stress that the proposed method leads to qualitative modifications of the topological properties of the system. Thus, in contrast to standard two-dimensional SSH arrays obtained by shifting the waveguides towards the periphery of the unit cell that support only one type of the corner state, in our square array twisting leads to simultaneous opening of two gaps, with corner states appearing in each of them. This is particularly surprising because for twisting scheme employed here, the complexity of the unit cell (the number of elements in it) does not increase in comparison with standard SSH array and remains the same for any twist angle. In clear contrast to super-SSH arrays [40], where new topological gaps open only due to the increase of the number of elements in the unit cell of the structure, in our case such gaps open because new types of coupling between waveguides arise as a result of twisting.

The method of HOTI construction proposed here is rather powerful and is not restricted exclusively to SSH-like geometries. It can be applied to periodic lattices of any symmetry, including honeycomb, Lieb, and kagome arrays with different number of waveguides in the unit cell and in each case it may considerably enrich the spectrum of the system, because twisting unavoidably leads to the appearance of additional couplings between different waveguides of the array, in contrast to traditional shifting of waveguides towards the periphery of the unit cell that changes only two types of coupling: intracell and intercell ones. Moreover, while such twisting preserves discrete rotational symmetry of the given array, it may involve different number of sites of the original structure (for example, in square array with $\mathcal{C}_4$ discrete rotational symmetry the twisting can be applied to nine sites instead of just four), and this is also expected to dramatically impact the spectrum of the structure and the number of emerging corner states. This highlights the potential of the method of twisting of waveguides in the unit cell for creation of novel topological phases, and for observation of new types of topological solitons with complex internal field distributions.

Paraxial propagation of the light beam in focusing nonlinear medium (along the $z$ axis) with transverse shallow modulation of the refractive index can be described by the nonlinear Schrödinger equation for the dimensionless light field amplitude $\psi$:

$$i\frac{\partial \psi}{\partial z} = -\frac{1}{2}\left(\frac{\partial^2 \psi}{\partial x^2} + \frac{\partial^2 \psi}{\partial y^2}\right) - |\psi|^2 \psi - \mathcal{R}(x,y)\psi, \quad (1)$$

where transverse coordinates $(x,y)$ are normalized to the characteristic scale $r_0 = 10\ \mu\mathrm{m}$; propagation distance $z$ is normalized to the diffraction length $kr_0^2$; $k=2\pi n/\lambda$ is the wavenumber; $n \approx 1.45$ is the unperturbed refractive index of the material; $\lambda = 800$ nm is the working wavelength; the intensity $|\psi|^2$ corresponds to $I = n|\psi|^2/k^2 r_0^2 n_2$, where $n_2$ is the nonlinear refractive index of the material (for instance, in fused silica $n_2 \approx 2.7 \times 10^{-20}\ \mathrm{m}^2/\mathrm{W}$). The function $\mathcal{R}(x,y)$ describes refractive index distribution inside the array that is composed from Gaussian waveguides $\mathcal{R}(x,y) = \sum_{l,m} \mathcal{Q}(x - x_m, y - y_l)$, where $\mathcal{Q}(x,y) = pe^{-(x^2+y^2)/a^2}$, of width $a = 0.5$ (5 $\mu$m), separated by the distance $d = 3$ (30 $\mu$m) and centered in the points $(x_m, y_l)$. Here waveguide depth $p = k^2 r_0^2 \delta n/n = 5$ is proportional to the refractive index contrast $\delta n \sim 5.6 \times 10^{-4}$ and selected such that all waveguides are single-mode.

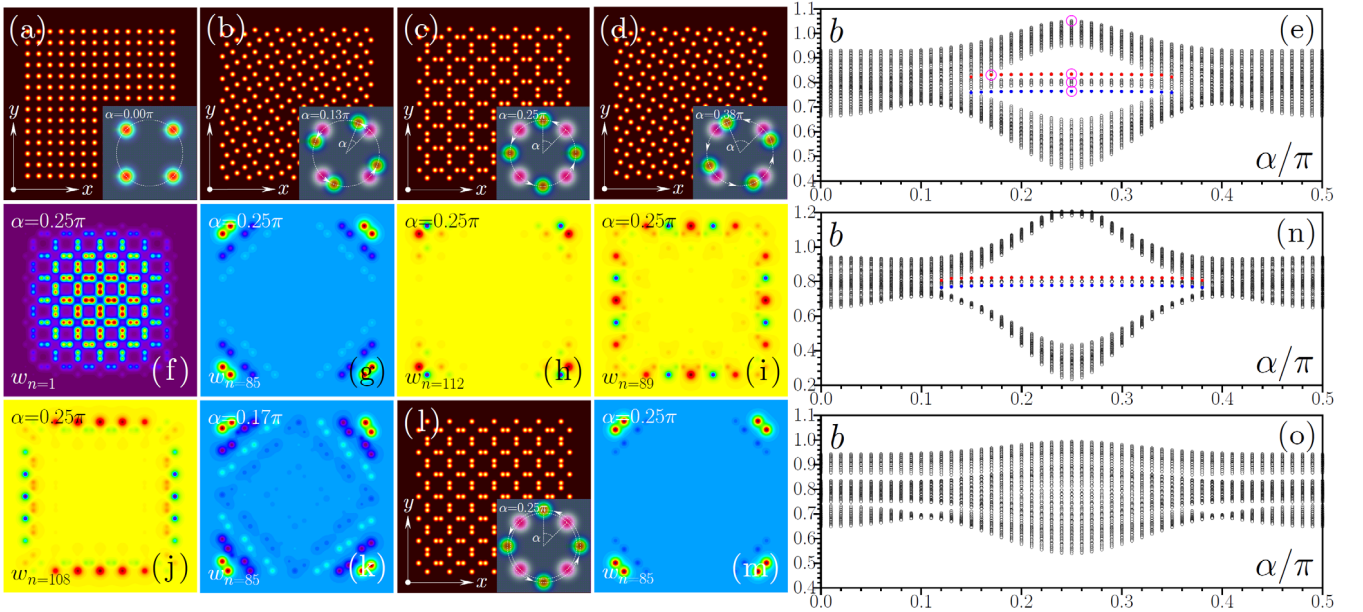

Fig. 1. (a)-(d) Examples of SSH waveguide arrays $\mathcal{R}$ with $7\times 7$ unit cells obtained upon twisting of waveguides inside each cell by the angle $\alpha$. Insets show waveguide positions inside unit cell before (red spots) and after (green spots) twist. (e) Propagation constant of linear eigenmodes of the array versus rotation angle $\alpha$. Open dots correspond to bulk modes, red and blue dots correspond to corner modes. (f)-(k) Examples of modes $w(x,y)$ corresponding to the magenta dots in linear spectrum (e). Twist angle and mode index are indicated on the plots. Profile of the array (l) and example of corner mode (m) in structure, where spacing between waveguides in the unit cell was increased by $\delta = 0.3$ prior to waveguide rotation by the angle $\alpha$. $b(\alpha)$ dependencies for arrays with $\delta = 0.3$ (n) and $\delta = -0.3$ (o). All eigenmodes and arrays are shown within $x,y \in [-26,+26]$ window.

Our method of construction of HOTI for the case of square array considered in this work is illustrated in Fig. 1(a)-1(d). The initial square array with period $d$ is depicted in Fig. 1(a). It is then decomposed into set of unit cells of width $2d$ with four waveguides per cell. The waveguides are then twisted by the same fixed angle $\alpha$ with respect to the axis passing through the center of each unit cell [see the insets in Fig. 1(b)-1(d) showing initial (magenta spots) and final (green spots) waveguide positions in the unit cell]. In the coordinate system with origin in the center of the unit cell, the coordinates $(x_m, y_l)$ of the waveguides change to $x'_m = x_m \cos\alpha - y_l \sin\alpha$, $y'_l = x_m \sin\alpha + y_l \cos\alpha$ after twist. Notice that for any angle $\alpha$ the structure maintains $\mathcal{C}_4$ discrete rotational symmetry, and it returns to the initial configuration (square array) for $\alpha = 0.5\pi$. Notice that twist does not change spacing between waveguides in the unit cell that is always equal to $d$, but it does affect spacing between waveguides in the neighboring cells. For instance, the spacing between two closest waveguides from neighboring cells varies as $s_1 = d[6 - 4(\cos\alpha + \sin\alpha)]^{1/2}$ and it becomes smaller than $d$ in the interval $0.095\pi < \alpha < 0.405\pi$. Within this interval one can expect the system to enter into topologically nontrivial regime because one of the intercell couplings becomes stronger than the intracell one. At the same time, the system is more complex than usual square SSH array (obtained by diagonal waveguide shifts rather than by twist), because the coupling between next-nearest waveguides from neighboring cells is non-negligible and must be taken into account too, since the distance $s_2 = d(5 - 4\cos\alpha)^{1/2}$ between them at $\alpha \leq 0.25\pi$ is comparable with $s_1$, at least for sufficiently small angles, and this may alter the actual interval of angles corresponding to topological phase. Notice that this structure is characterized by simultaneous appearance in topological phase of two well-separated waveguides in each corner, visible in Fig. 1(c) at $\alpha = 0.25\pi$, in contrast to usual SSH array, where only one such waveguide appears in the corner. One therefore may expect that the system considered here will have more complex modal spectrum.

To calculate it, we first omit nonlinearity in Eq. (1) and search for eigenmodes of the system in the form $\psi(x,y,z) = w_n(x,y)e^{ib_nz}$, where $b_n$ is the propagation constant of the mode with index $n$. We sort eigenmodes such that their propagation constants decrease with $n$. Further we consider the array $\mathcal{R}$ with $7\times7$ unit cells, total number of waveguides is $196$ [Fig. 1(a)-1(d)]. The dependence of propagation constants of all eigenmodes of the array on twist angle $\alpha$ is presented in Fig. 1(e). One can see that in this system two topological gaps simultaneously open in the spectrum for $0.15\pi \leq \alpha \leq 0.35\pi$. While modes in the bands (open dots) are delocalized and extend over entire array [Fig. 1(f)], the modes in two gaps corresponding to red and blue dots are topological modes localized in four corners of the structure [Fig. 1(g) and 1(h)]. Because the structure is $\mathcal{C}_4$ symmetric, such modes emerge simultaneously in all four corners (i.e. each red and blue dot in spectrum corresponds to four nearly degenerate states – linear combinations of modes residing in different corners). Corner modes from different gaps have different internal structure. In the mode from top gap two spots in corner waveguides are in-phase [Fig. 1(g)], while in the mode from bottom gap these spots are out-of-phase [Fig. 1(h)]. Two gaps are separated by the band of edge states, the examples of such states are shown in Fig. 1(i) and 1(j). Notice that corner modes are most localized at $\alpha = 0.25\pi$, when the widths of two gaps are largest. Corner modes extend and become notably asymmetric with respect to diagonal of the array as twist angle approaches $\alpha \approx 0.15\pi$ or $0.35\pi$, when gaps close [Fig. 1(k)]. It should be mentioned, that the interval of twist angles, where the system is topological can be increased if prior to twisting the waveguides, they are shifted along the diagonals towards the corners of the unit cell by a distance $2^{-1/2}\delta$ (in this case, the intracell waveguide spacing increases to $d + \delta$). The example of such array for $\delta = 0.3$ is shown in Fig. 1(l), while its spectrum as a function of twist angle $\alpha$ is illustrated in Fig. 1(n). The widths of the gaps in this case also notably increase, leading to more pronounced localization of corner modes [Fig. 1(m)]. In contrast, when shift $\delta$ of waveguides prior to twisting is negative (leading to reduction of intracell spacing), the gaps may disappear for all values of $\alpha$ [Fig. 1(o)] and corner states disappear too. Further upon analysis of solitons we concentrate on the $\delta = 0$ case.

Topological solitons bifurcate under the action of nonlinearity from localized corner modes. Such solitons can be found from Eq. (1) in the form $\psi = we^{ibz}$, where real-valued function $w(x,y)$ satisfies the equation

$$\frac{1}{2}\left(\frac{\partial^2 w}{\partial x^2} + \frac{\partial^2 w}{\partial y^2}\right) + w^3 + \mathcal{R}w - bw = 0, \quad (2)$$

that was solved using Newton method. Soliton families in the form of dependencies of power $U = \iint |\psi|^2 \, dxdy$ on propagation constant $b$ are presented in Fig. 2(a) for the case $\alpha = 0.25\pi$. We consider here only solitons emerging in top or bottom topological gaps [white regions in Fig. 2(a)] from two types of corner modes with in-phase and out-of-phase spots discussed in Fig. 1. In the bifurcation point propagation constant of soliton coincides with that of linear corner mode, while its power $U$ vanishes (i.e. such states are thresholdless in contrast to usual two-dimensional lattice solitons [72]). The amplitude and power of soliton grow as one moves away from bifurcation point and for sufficiently large $U$ nonlinearity drives soliton into the band [gray regions in Fig. 2(a)]. Thus the family of the out-of-phase corner solitons that are localized inside the bottom gap [mode 1 in Fig. 2(c)] enter the band of edge states, where coupling with such states occurs, that results in the appearance of multiple peaks on the edge of the array [mode 2 in Fig. 2(d)]. In contrast, in-phase corner soliton emerging in the top gap and localized insider this gap [mode 5 in Fig. 2(g)], enters for sufficiently large $U$ into band of bulk states, that results in strong expansion into the depth of array [mode 6 in Fig. 2(h)]. Notice that one can also find a "continuation" of the out-of-phase corner soliton family in the top gap [see modes 3 and 4 in Fig. 2(e), (f)]. Close inspection of corresponding solutions show that they differ by structure of tails from solitons in the bottom gap. In addition, the upper branch of such solutions (mode 4) in the top gap is characterized by the presence of additional intensity maxima in the unit cells adjacent to the corner one. In fact, multiple soliton families visible in Fig. 2(a) emerge due to avoided crossings with nonlinear families emanating from different edge states [left gray region in Fig. 2(a)]. As mentioned above, when the family of corner solitons enters the band, its slope abruptly increases and this occurs such as to avoid direct crossing of this family with any nonlinear family emanating from edge modes. The same occurs for the family of out-of-phase corner solitons residing in the upper gap, that looks like a continuation of the family in lower gap – but this particular family just turns up near the band of the edge states, acquiring nonzero power threshold. Figure 2(a) shows only these two outermost families, and do not show multiple families emanating from edge states. One can also see that abrupt change of slope for corner soliton families $U(b)$ occurs in different points within the allowed band. This depends on where exactly inside this band one observes the avoided crossing with nonlinear edge state family and what is the symmetry of states from corresponding edge state family. We believe that the symmetry of such states should closely match the symmetry of corner states, only in this case the efficient interaction between them would be possible. One can see that nonlinearity offers an efficient tool to control localization degree and shapes of topological states.

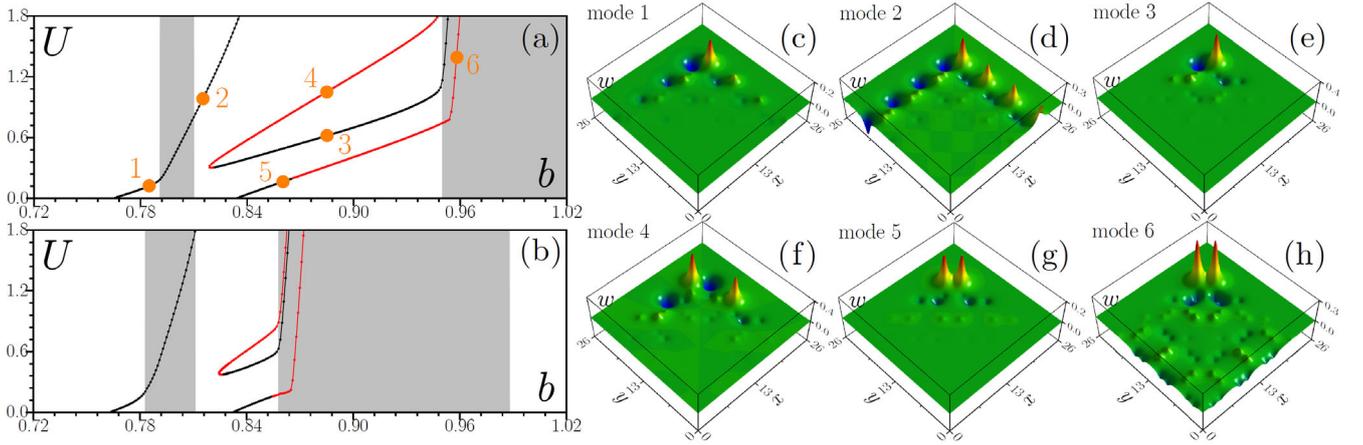

Fig. 2. $U(b)$ dependencies illustrating families of corner solitons bifurcating from the out-of-phase and in-phase topological corner states at $\alpha=0.25\pi$ (a) and $\alpha=0.18\pi$ (b). Black branches correspond to stable solitons, while red branches correspond to unstable solitons. Left and right gray regions show bands of edge and bulk states, respectively. Representative soliton profiles at $b=0.785$ (c), $0.815$ (d), $0.885$ (e),(f), $0.860$ (g), and $0.958$ (h) correspond to orange dots in (a) at $\alpha=0.25\pi$. In all cases $\delta=0$.

Similar soliton families can be obtained for other twist angles $\alpha$, see for example Fig. 2(b) for $\alpha=0.18\pi$. The widths of both gaps are smaller in this case, hence solitons enter into the bands at lower power levels $U$. Field modulus distributions of solitons at $\alpha=0.18\pi$ become asymmetric with respect to diagonal of the array.

The very fact of simultaneous appearance of two linear topological corner modes in the spectrum does not guarantee yet that both soliton branches bifurcating from such linear states can be stable in the presence of focusing nonlinearity (because such states have different phase structure). Indeed, different symmetry of corner solitons forming in two gaps is manifested in different stability properties. To analyze stability of the obtained solutions we perturb them with small-scale noise and propagate them in the frames of Eq. (1) up to $z=10^4$. In the case when soliton exhibits only small-amplitude oscillations due to initial perturbation, it is considered stable, while if perturbations grow causing changes in the amplitude and internal structure of solution, the latter is considered unstable. Black branches in Fig. 2(a) and 2(b) correspond to stable solitons, while red branches correspond to unstable states. One can see that the entire branch of the out-of-phase corner solitons is stable. Example of stable propagation of such soliton is presented in Fig. 3(a), where we show peak amplitude $\max|\psi|$ of solution versus propagation distance and field modulus distributions at different distances. Remarkably, this branch of solutions remains stable even when it enters into the band of edge states and couples with them.

We also found that lower branch of the "continuation" of the out-of-phase family in the top gap is also dynamically stable, except for a very narrow region of propagation constants, where it connects with the upper branch [see branches with dots 3 and 4 in Fig. 2(a)]. The example of stable propagation of mode 3 from Fig. 2(a) belonging to this lower branch is illustrated in Fig. 3(b). In contrast, the upper branch of the out-of-phase solitons in top gap is entirely unstable. As concerns the family of the in-phase corner solitons emerging in the top gap, it can be stable only in the narrow interval of propagation constants adjacent to the bifurcation point from linear corner state. Even though the stability interval for this state is narrow, it shows that several stable corner soliton families can coexist in higher-order topological insulators. To the best of our knowledge, this coexistence has never been reported for HOTIs. Increasing $U$ leads to destabilization of this family of solutions. Typical example of instability development is presented in Fig. 3(c). It is accompanied by strong oscillations of two intensity maxima in corner waveguides and pronounced radiation into the bulk, particularly well visible at $z=240$. Interestingly, radiation into the bulk leads to gradual decrease of power concentrated around the corner and this state may eventually enter into stability domain, after shedding away considerable fraction of initial power. Similar structure of stability and instability domains is obtained for other angles, including $\alpha=0.18\pi$ [Fig. 2(b)].

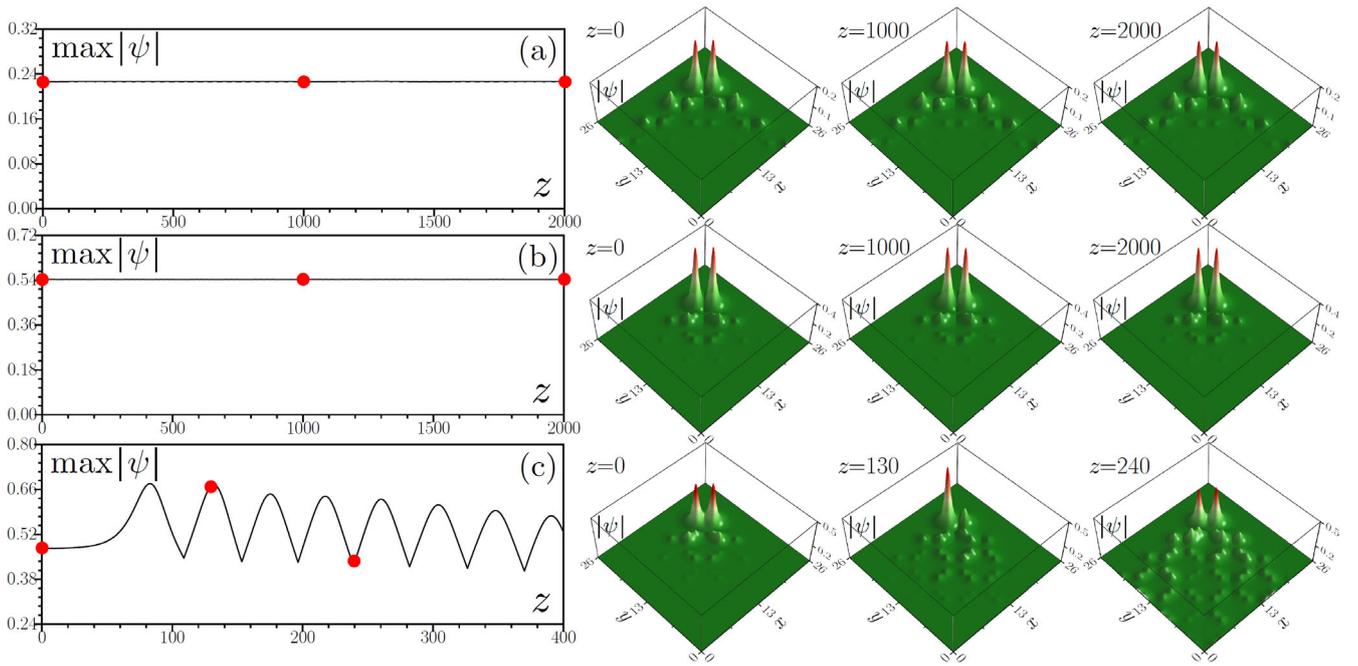

Fig. 3. Stable propagation of corner solitons with (a) $b=0.785$ (mode 1 in Fig. 2) and (b) $b=0.885$ (mode 3 in Fig. 2) bifurcating from the out-of-phase topological corner state. (c) Decay of the unstable corner soliton with $b=0.920$ bifurcating from the in-phase corner state. Field modulus distributions shown at different distances correspond to the red dots in $\max|\psi|$ vs $z$ dependence. In all cases $\alpha=0.25\pi$, $\delta=0$.

Summarizing, I have shown that twisting of the waveguides in the unit cell of two-dimensional periodic structure can be used to construct HOTIs supporting topological corner states in certain interval of twist angles, where two forbidden topological gaps open in the linear spectrum. The method proposed here can be extended to other types of periodic lattices, in particular, to honeycomb, Lieb and kagome ones, whose spectrum can be also dramatically enriched due to waveguide twisting, particularly because twisting can be applied to groups of waveguides with different number of elements. This method therefore opens a new avenue for construction of topologically nontrivial phases with different, much richer structure of the modal spectrum as compared to structures obtained using standard approach, when waveguides are shifted only along the diagonal or perpendicularly to the border of the unit cell. The structure proposed here allows to obtain two different coexisting types of two-dimensional corner solitons from different gaps, the possibility that was not illustrated in HOTIs before, to the best of our knowledge. It should be also mentioned that if the twist angle $\alpha$ of waveguides in this system were linearly increasing with propagation distance $z$, resulting in helical waveguide trajectories, one could potentially realize on the basis of such structure a new type of Floquet HOTI hosting periodically oscillating Floquet corner modes of various types.

**Acknowledgements:** Y.V.K. acknowledges funding by the Russian Science Foundation grant 24-12-00167 and partially by the research project FFUU-2024-0003 of the Institute of Spectroscopy of the Russian Academy of Sciences.